\documentclass[12pt]{iopart}
\usepackage{epsfig}
\usepackage{latexsym}
\usepackage{graphicx}
\usepackage[usenames,dvipsnames]{color}
\def\lsim{\raise0.3ex\hbox{$<$\kern-0.75em\raise-1.1ex\hbox{$\sim$}}}
\def\gsim{\raise0.3ex\hbox{$>$\kern-0.75em\raise-1.1ex\hbox{$\sim$}}}

\def\mean#1{\left<#1\right>}
\def\Journal#1#2#3#4{{#1}{\bf #2}, #3 (#4)}

\def\NPA{{Nucl. Phys. A}}
\def\NPB{{Nucl. Phys. B}}
\def\PLB{{Phys. Lett. B}}

\def\PRL{Phys. Rev. Lett.\ }
\def\PRD{{Phys. Rev. D}}
\def\PRC{{Phys. Rev. C}}

\begin{document}

\title{Measuring Parton Energy Loss at RHIC compared to LHC}

\author{M.~J.~Tannenbaum\footnote{Supported by the U.S. Department of Energy, Contract No. DE-AC02-98CH1-886.}, PHENIX Collaboration}

\address{Physics Department, Brookhaven National Laboratory, Upton, NY 11973-5000, USA}
\ead{mjt@bnl.gov}

\begin{abstract}
The method of measuring $\hat{x}_h=\hat{p}_{Ta}/\hat{p}_{Tt}$, the ratio of the away-parton transverse momentum,  $\hat{p}_{T_a}$, to the trigger-parton transverse momentum, $\hat{p}_{T_t}$, using two-particle correlations at RHIC, will be reviewed. This measurement is simply related to the two new variables introduced at LHC for the di-jet fractional transverse momentum imbalance: ATLAS $A_J=(\hat{p}_{Tt}-\hat{p}_{Ta})/(\hat{p}_{Tt}+\hat{p}_{Ta})= (1-\hat{x}_h)/(1+\hat{x}_h)$; and CMS $\mean{(\hat{p}_{Tt}-\hat{p}_{Ta})/\hat{p}_{Tt}}= \mean{1-\hat{x}_h}$. Results from two-particle correlations at RHIC for $\hat{x}_h$ in p-p and A+A collisions will be reviewed and new results will be presented and compared to LHC results. The importance of comparing any effect in A+A collisions to the same effect effect in p-p collisions will be illustrated and emphasized. 

\end{abstract}
\section{Introduction}
  In 1998, at the QCD workshop in Paris, Rolf Baier asked me whether jets could be measured in Au+Au collisions because he had a prediction of a QCD medium-effect (energy loss via  soft gluon radiation induced by multiple scattering~\cite{BDPS} on color-charged partons traversing a hot-dense-medium composed of screened color-charges~\cite{BaierQCD98}). I told him~\cite{MJTQCD98} that there was a general consensus~\cite{Strasbourg} that for Au+Au central collisions at $\sqrt{s_{NN}}=200$ GeV, leading particles are the only way to study jets, because in one unit of the nominal jet-finding cone,  $\Delta r=\sqrt{(\Delta\eta)^2 + (\Delta\phi)^2}$, there is an estimated $\pi\Delta r^2\times{1\over {2\pi}} {dE_T\over{d\eta}}\sim 375$ GeV of energy !(!) The good news was that hard-scattering in p-p collisions was originally observed by the method of leading particles and that these techniques could be used to study hard-scattering and jets in Au+Au collisions~\cite{MJTEPS04}.    
  \section{Hard scattering via single particle inclusive and two-particle correlation measurements}
Single particle inclusive and two-particle correlation measurements of hard-scattering have provided a wealth of discoveries at RHIC. Due to the steeply falling power-law invariant transverse momentum spectrum of the scattered parton, $\hat{p}^{-n}_{T_t}$, the inclusive single particle (e.g. $\pi^0$) $p_{T_t}$ spectrum from jet fragmentation is dominated by fragments with large $z_{\rm trig}$, where $z_{\rm trig}=p_{T_t}/\hat{p}_{T_t}$ is the fragmentation variable, and exponential fragmentation $D^{\pi^0}_q(z)\sim e^{-bz}$ is assumed. This gives rise to several effects which allow precision measurements of hard scattering to be made using single inclusive particle spectra and two particle correlations~\cite{egMJTPoS06,MJTPoS06}.
 
 	The prevailing opinion from the 1970's until quite recently was that although the inclusive single particle (e.g. $\pi^0$) spectrum from jet fragmentation is dominated by trigger fragments with large $\mean{z_{\rm trig}}\sim 0.6-0.8$, the away-jets should be unbiased and would measure the fragmentation function, once the correction is made for $\mean{z_{\rm trig}}$ and the fact that the jets don't exactly balance $p_T$ due to the $k_T$ smearing effect~\cite{FFF}.  Two-particle correlations with trigger $p_{T_t}$, are analyzed in terms of the two variables: $p_{\rm out}=p_{T} \sin(\Delta\phi)$, the out-of-plane transverse momentum of an associated track with $p_T$;  
 and $x_E$, where:\\
 \begin{center}\vspace*{-1.5pc} 
\begin{minipage}[b]{0.45\linewidth}	
$$x_E=\frac{-\vec{p}_{T}\cdot \vec{p}_{Tt}}{|p_{Tt}|^2}=\frac{-p_{T} \cos(\Delta\phi)}{p_{Tt}}\simeq \frac {z}{z_{\rm trig}}$$ 
\vspace*{0.06in}
\end{minipage}
\hspace*{0.01\linewidth}
\begin{minipage}[b]{0.450\linewidth}
\vspace*{0.06in}
\includegraphics[scale=0.6]{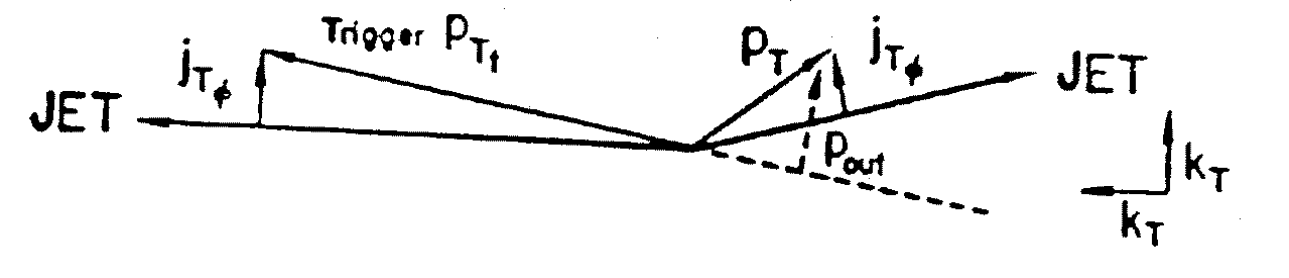}
\vspace*{-0.12in}
\label{fig:mjt-poutxe}
\end{minipage}\vspace*{-2.0pc}
\end{center}
$z_{\rm trig}\simeq p_{Tt}/p_{T{\rm jet}}$ is the fragmentation variable of the trigger jet, and $z$ is the fragmentation variable of the away jet. 

	However, in 2006, it was found by explicit calculation that this is not true~\cite{ppg029,egMJTPoS06,MJTPoS06}. The shape of the $p_{T_a}$ spectrum of fragments (from the away-side parton with $\hat{p}_{T_a}$), given a trigger particle with $p_{T_t}$ (from a trigger-side parton with $\hat{p}_{T_t}$), is not sensitive to the shape of the fragmentation function ($b$), but measures the ratio of $\hat{p}_{T_a}$ of the away-parton to $\hat{p}_{T_t}$ of the trigger-parton and depends only on the same power $n$ as the invariant single particle spectrum:  
		     \begin{equation}
\left.{dP_{p_{T_a}} \over dx_E}\right|_{p_{T_t}}\approx {\mean{m}(n-1)}{1\over\hat{x}_h} {1\over
{(1+ {x_E \over{\hat{x}_h}})^{n}}} \, \qquad . 
\label{eq:condxe2}
\end{equation}
This equation gives a simple relationship between the ratio, $x_E\approx p_{T_a}/p_{T_t}\equiv z_T$, of the transverse momenta of the away-side particle to the trigger particle, and the ratio of the transverse momenta of the away-jet to the trigger-jet, $\hat{x}_{h}=\hat{p}_{T_a}/\hat{p}_{T_t}$. The only dependence on the fragmentation function is in the mean multiplicity $\mean{m}$ of jet fragments. This functional form was shown previously~\cite{ppg029, ppg095} (and with the present data, see below) to describe the $\pi^0$ triggered $x_E$ distribution in p-p collisions and is based only on the following simplifying assumptions:  the hadron fragment is assumed to be collinear 
with the parton direction; the underlying fragmentation functions ($D(z)$) are assumed to be exponential; and for a given $p_{T_t}$, $\hat{x}_h$ is taken to be constant as a function of $x_E$ over the range of interest.  The key issue with Eq.~\ref{eq:condxe2} is that it is independent of the slope of an exponential fragmentation function, and only depends on the detected mean multiplicity $\mean{m}$ of the jet, the power, $n$, of the inclusive $p_{T_t}$ spectrum and the ratio of the away jet to the trigger jet transverse momenta, $\hat{x}_h$. 

\section{Fits to PHENIX $\pi^0$-h correlations}
\begin{figure}[!h]
   \begin{center}
\includegraphics[width=0.66\linewidth]{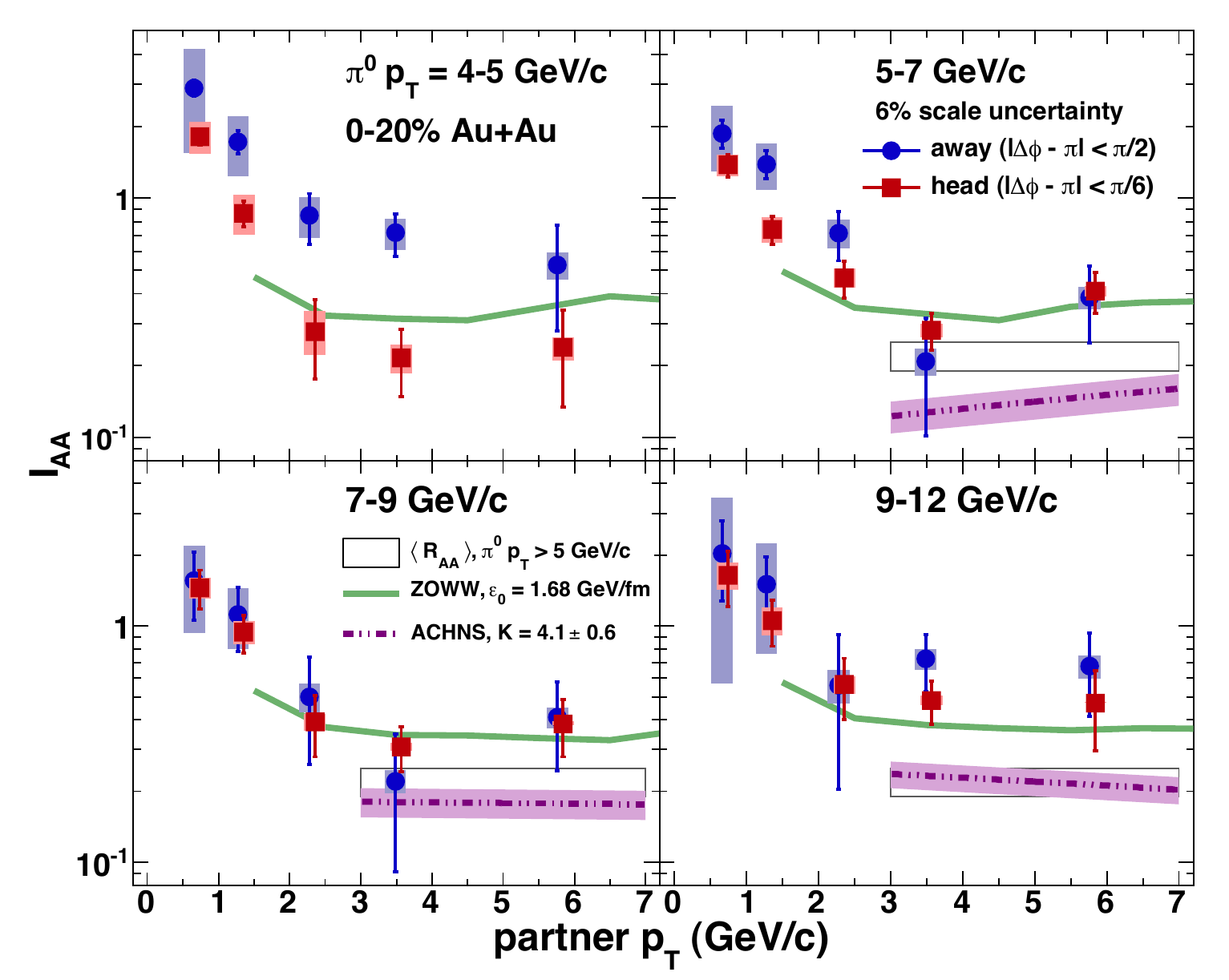}
\end{center}\vspace*{-1pc}
\caption[]{\footnotesize Away-side $I_{AA}$~\cite{ppg106} for a narrow ``head'' $|\Delta\phi -\pi|<\pi/6$ selection (solid squares) and the entire away-side, $|\Delta\phi -\pi|<\pi/2$ (solid circles) as a function of partner momentum $p_{T_a}$ for various trigger momenta $p_{T_t}$. Only the head region was used for the present analysis.\label{fig:ppg106IAA}}
\end{figure}

 The two-particle correlation distributions from $\pi^0$ triggers in four intervals of $p_{T_t}$, 4-5, 5-7, 7-9 and 9-12 GeV/c, with charged hadrons in a fixed range of of associated transverse momenta, $p_{T_a}\approx 0.7, 1.3, 2.3, 3.5, 5.8$ GeV/c were recently published by PHENIX~\cite{ppg106} in terms of the ratio of A+A to p-p collisions, $I_{AA}(p_{T_a})|_{p_{T_t}}=\left.\frac{dP^{AA}/dp_{Ta}}{dP^{pp}/dp_{Ta}}\right|_{p_{T_t}}$ (see Fig.~\ref{fig:ppg106IAA}). 

We now analyze these distributions separately for p-p and Au+Au collisions, with the statistical error and the larger of the $\pm$ systematic errors of the data points added in quadrature. The p-p and Au+Au distributions in $z_T=p_{T_a}/p_{T_t}$ were fit to the formula~\cite{ppg029}:
    \begin{equation}
\left.{dP_{\pi} \over dz_T}\right|_{p_{T_t}}  = {N\,(n-1)}{1\over\hat{x}_h} {1\over {(1+ {z_T \over{\hat{x}_h}})^{n}}} \,  
\qquad ,  
\label{eq:condxe2N}
\end{equation}   
with a fixed value of $n=8.10$ ($\pm0.05$) as previously determined~\cite{ppg080} , where $n$ is the power-law of the inclusive $\pi^0$ spectrum and is observed to be the same in p-p and Au+Au collisions in the $p_{T_t}$ range of interest.  The fitted value for $N$ is the integral of the $z_T$ distribution which equals $\mean{m}$, the mean multiplicity of the away jet in the PHENIX detector acceptance, and $\hat{x}_h\equiv \hat{p}_{T_a}/\hat{p}_{T_t}$ is the ratio of the away jet to the trigger jet transverse momenta. 

Fits were performed for the p-p spectra; and also for the Au+Au spectra at two centralities: 0-20\% and 20-40\% upper-percentiles.  The parameters of the p-p distribution, $\hat{x}^{pp}_h$ and $N_{pp}$, are determined by fits of Eq.~\ref{eq:condxe2N} to the p-p data for the four intervals of $p_{T_t}$; and the parameters $\hat{x}^{AA}_h$ and $N_{AA}$ are determined from the fits to the Au+Au distributions. The fits were performed only for the narrower ``head'' region, $|\Delta\phi -\pi|<\pi/6$. It should be noted that in Fig.~\ref{fig:ppg106IAA}, there is no difference in the results ($I_{AA}$) for the full away side and the head region, for $p_{T_t}\geq 7$ GeV/c, because the non-jet background becomes sufficiently small so that the ``shoulder''~\cite{PXPRC77}, now known to be due to a $v_3$ background modulation~\cite{PXv3} for which no correction has been applied in this data, contributes negligibly to the away-side yield.    

\section{Results of the fits}
\begin{figure}[!bh]
\begin{center}
a)\includegraphics[width=0.44\linewidth]{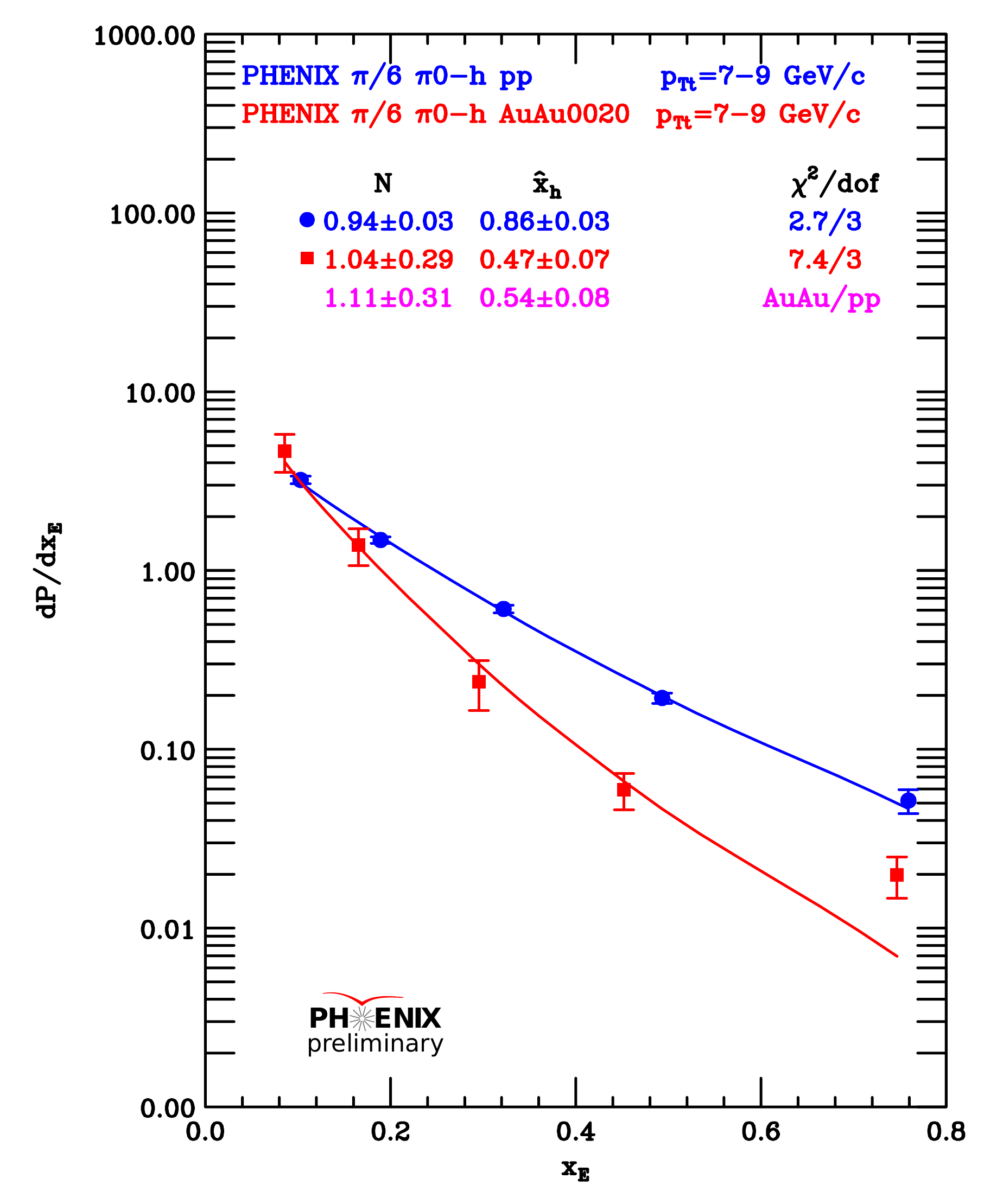}
b)\includegraphics[width=0.44\linewidth]{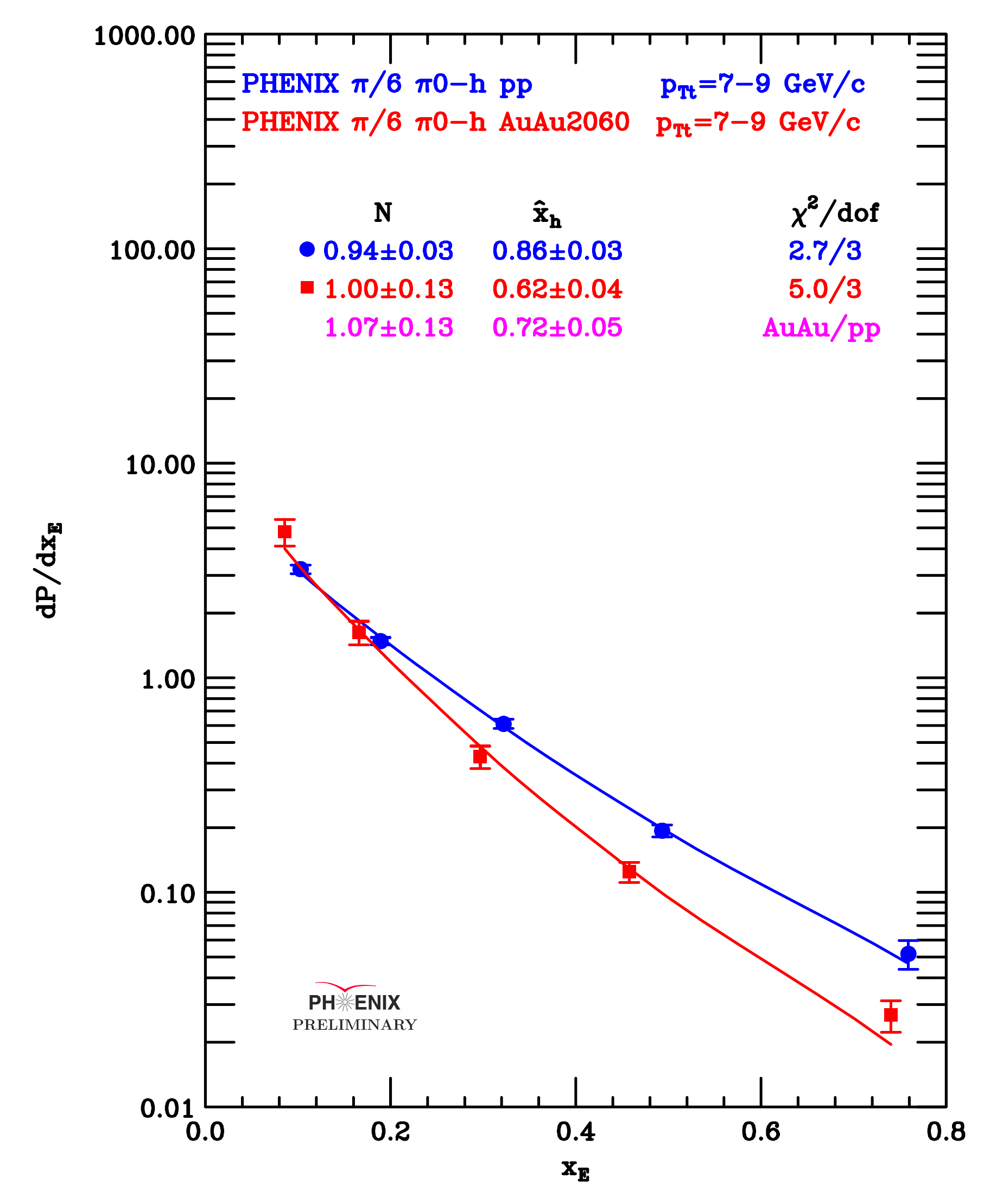}
\end{center}\vspace*{-0.20in}
\caption[]{ p-p (blue circles) and AuAu  (red squares) $z_T=p_{T_a}/\mean{p_{T_t}}$ distributions for $p_{T_t}=7-9$ GeV/c ($\mean{p_{T_t}}=7.71$ GeV/c), together with fits to Eq.~\ref{eq:condxe2N} p-p (solid blue line), AuAu (solid red line) with parameters indicated: a) 00-20\% centrality, b) 20--60\% centrality. The ratios of the fitted parameters for AuAu/pp are also given. }
\label{fig:AuAupp79}
\end{figure}

Examples of the fits for $7<p_{T_t}<9$ GeV/c for p-p collisions and Au+Au 0--20\% and 20--60\% are shown in Figs.~\ref{fig:AuAupp79}a and b, respectively. 
The results for the fitted parameters are shown on the figures. In general the values of $\hat{x}^{pp}_h$ do not equal 1 but range between $0.8<\hat{x}^{pp}_h<1.0$ due to $k_T$ smearing and the range of $z_T$ covered. For the fixed range of associated $p_{T_a}$ $0.7-5.8$ GeV/c, the lowest $p_{T_t}=4-5$ GeV/c trigger provides the most balanced same and away side jets, with $\hat{x}_h\approx1.0$, while as $p_{T_t}$ increases up to 9--12 GeV/c, for the fixed range of $p_{T_a}$, the jets become unbalanced towards the trigger side in p-p collisions due to $k_T$ smearing. Thus, in the present data, the $p_{T_t}$ and $z_T$ ranges are identical for the p-p and Au+Au comparison. Furthermore, in order to take account of the imbalance ($\hat{x}^{pp}_h <1$) observed in the p-p data, the ratio $\hat{x}_h^{AA}/\hat{x}_h^{pp}$ is taken as the measure of the energy of the away jet relative to the trigger jet in A+A compared to p-p collisions. 

It is important to note that the away jet energy fraction in AuAu relative to p-p, $\hat{x}_h^{AA}/\hat{x}_h^{pp}=0.47/0.86=0.54\pm0.08$ in Fig.~\ref{fig:AuAupp79}a,  is significantly less than 1, indicating energy loss of the away jet in the medium. Also since the away-jet may suffer different energy losses for a given trigger jet $\hat{p}_{T_t}$ due to variations in the path-length through the medium, $\hat{x}_h^{AA}$ should be understood as $\mean{\hat{x}_h^{AA}}$. 
\section{LHC Results}
\begin{figure}[!h]
\begin{center}
\includegraphics[width=0.75\linewidth]{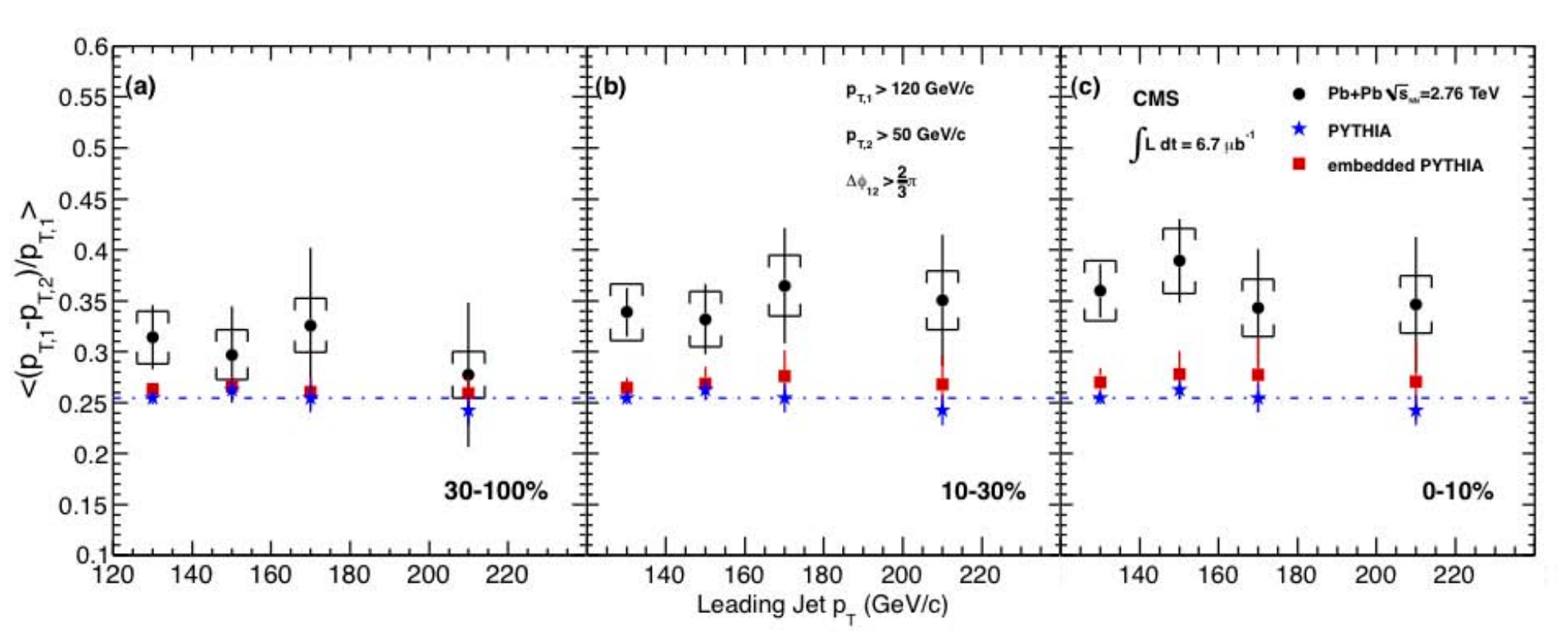}
\end{center}\vspace*{-1pc}
\caption[]{\footnotesize CMS~\cite{CMS} plot of $\mean{1-p_{t,2}/p_{t,1}}$, the fractional jet imbalance, as a function of $p_{T,1}$ for 3 centralities in p-p and Pb+Pb collisions. }
\label{fig:LHC}
\end{figure}

In very exciting first results from the LHC heavy ion program, ATLAS~\cite{ATLAS} observed dijet events in Pb+Pb central collisions at $\sqrt{s_{\rm NN}}=2.76$ TeV with a large energy asymmetry which they characterized by a new quantity $A_J=(1-\hat{x}_h^{AA})/(1+\hat{x}_h^{AA})$.  
Shortly thereafter,  CMS~\cite{CMS} presented a plot of $\mean{1-p_{t,2}/p_{t,1}}=1-\mean{\hat{x}_h^{AA}}$, the fractional jet imbalance as a function of $E_{T1}$ up to 200--220 GeV with a cut $E_{T2}\geq 50$ GeV  (Fig.~\ref{fig:LHC}). If there were no cuts on the p-p jets used in this measurement, then this variable should be identical to the one we call $1-\hat{x}_h^{AA}/\hat{x}_h^{pp}$, the away-parton fractional energy loss (or imbalance) in A+A relative to p-p.  
However, due to the cut used in the CMS data, the sample of di-jets in p-p used to compare with A+A  suffers from a large imbalance of 0.25, independent of $E_{T1}$ (Fig.~\ref{fig:LHC}). We correct this by calculating $\hat{x}_h^{AA}$ and $\hat{x}_h^{pp}$ for CMS from their 
given values of $1-\hat{x}_h^{AA}$ and $1-\hat{x}_h^{pp}$ and then correcting to $1-\hat{x}_h^{AA}/\hat{x}_h^{pp}$. For instance, in Fig.~\ref{fig:LHC}c for $E_{T1}=130$ GeV, $\mean{1-\hat{x}_h^{pp}}=0.255$ (i.e. $\mean{\hat{x}_h^{pp}}=0.745$), while $\mean{1-\hat{x}_h^{AA}}=0.36$ (i.e. $\mean{\hat{x}_h^{AA}}=0.64$), so that $1-\mean{\hat{x}_h^{AA}}/\mean{\hat{x}_h^{pp}}=1-(0.64/0.745)=0.141$. 

   \begin{figure}[!h]
\begin{center}
\includegraphics[width=0.8\linewidth]{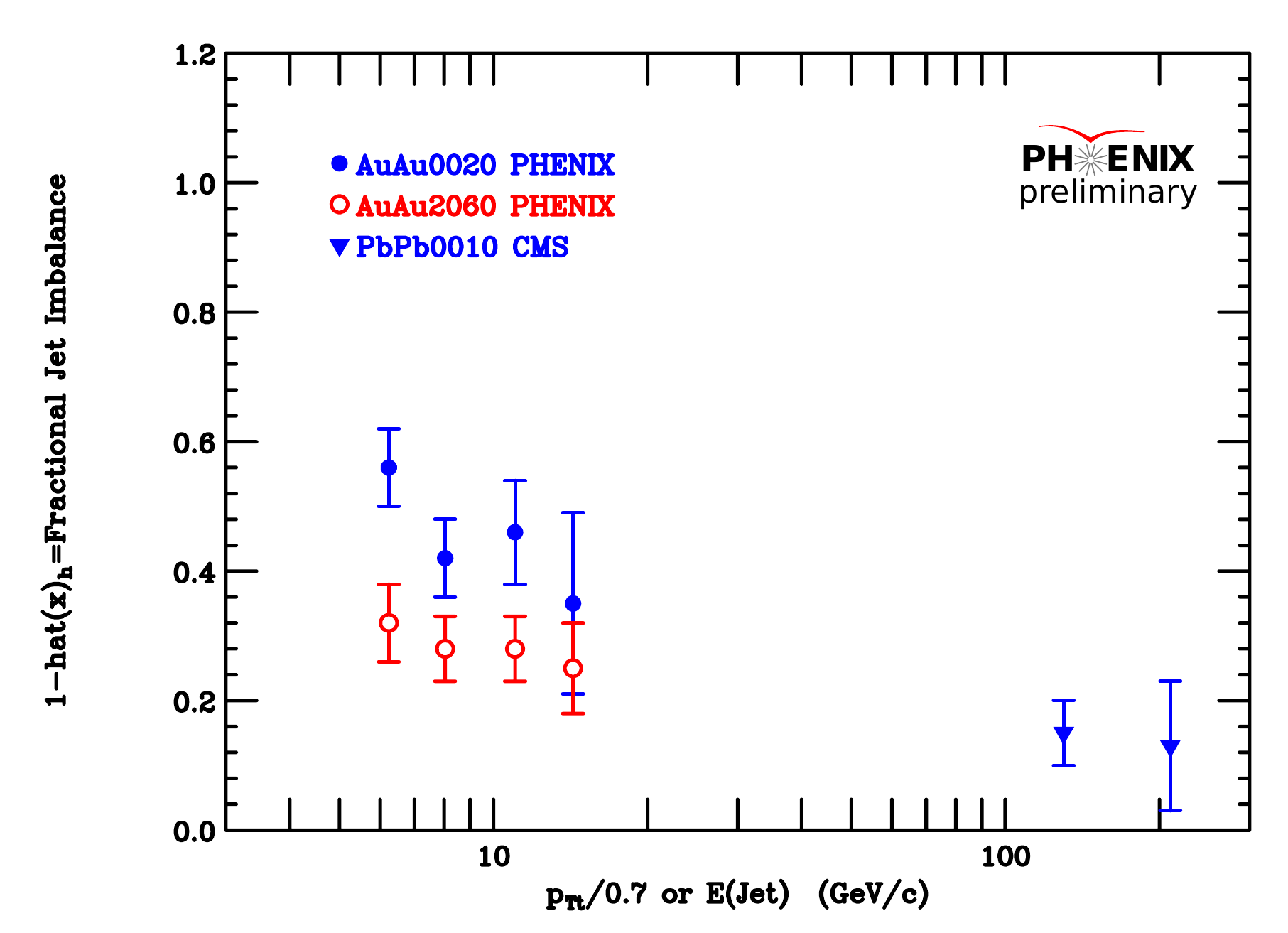}
\end{center}\vspace*{-0.30in}
\caption[]{\footnotesize Away-jet fractional imbalance or energy loss in A+A relative to p-p, $1-\hat{x}_h$, as a function of $p_{T_t}/0.7$ for PHENIX and E(Jet) for CMS, with centralities indicated.}
\label{fig:PXLHCxh}\vspace*{-1pc}
\end{figure}

The corrected points are shown together with the PHENIX data for $1-\hat{x}_h^{AA}/\hat{x}_h^{pp}$, which we denote for simplicity $\mean{ 1-\hat{x}_h}$, the observed fractional jet imbalance in A+A relative to p-p (Fig.~\ref{fig:PXLHCxh}). Of course the CMS result is directly measured with jets, while the PHENIX value is deduced from the fragments of the dijets using a few simple  assumptions, as noted above. The PHENIX data are plotted at the presumed mean trigger parton transverse momentum $\mean{\hat{p}_{T_t}}=p_{T_t}/\mean{z_{\rm trig}}$, where the average fragmentation fraction of the trigger particle, $\mean{z_{\rm trig}}\approx 0.7$, was derived in Ref.~\cite{ppg029}.  
There is a clear difference in fractional jet imbalance in going from RHIC to LHC in central collisions---the jet-imbalance or fractional energy loss is much smaller at LHC. This is different from the first impression~\cite{ATLAS}. Also at RHIC, there is less fractional energy loss or jet imbalance in less central collisions.  

The large difference in fractional jet imbalance between RHIC and LHC c.m. energies could be due to the difference in jet $\hat{p}_{T_t}$ between RHIC ($\sim 20$ GeV/c) and LHC ($\sim 200$ GeV/c), the difference in $n$ for the different $\sqrt{s}$, or to a difference in the properties of the medium. Future measurements will need to sort out these issues by extending both the RHIC and LHC measurements to overlapping regions of $p_T$.

\end{document}